\documentclass[12pt,a4paper]{article}
\usepackage{latexsym}
\font\mybb=msbm10 at 12pt
\def\bb#1{\hbox{\mybb#1}}
\def\Z {\bb{Z}}

\begin{document}

\begin{flushright}
CERN--TH/96--214\\
{\bf hep-th/9608044}\\
August $8$th, $1996$
\end{flushright}

\begin{center}


{\Large {\bf Massive and Massless Supersymmetric\\ Black Holes}},

\vspace{.3cm}

{\small Contribution to the Proceedings of the {\it Trieste Spring
Workshop on String Theory, Gauge Theory and Quantum Gravity}, held in
the I.C.T.P., Trieste, (Italy), the 28th and 29th of March 1996.}

\vspace{.9cm}

{\large
{\bf Tom\'as Ort\'{\i}n}
\footnote{E-mail address: {\tt Tomas.Ortin@cern.ch}}\\
\vspace{.4cm}
{\it C.E.R.N.~Theory Division}\\
{\it CH--1211, Gen\`eve 23, Switzerland}\\
}

\vspace{.8cm}


{\bf Abstract}

\end{center}

\begin{quotation}

\small

We review some recent work on the existence and classification of
extreme black-hole-type solutions in $N=8$ supergravity. For the black
holes considered (those that are also solutions of $N=4$ supergravity
and of the Einstein-Maxwell-dilaton theory with coupling $a$) a
complete classification is achieved: the only possible values of $a$
are $\sqrt{3},1,1/\sqrt{3},0$. Up to U~duality transformations there
is only one solution for each of those values.  The exception is $a=0$
for which an additional extreme but non-supersymmetric
Reissner-Nordstr\"om black hole solution exists.

We also study the so-called massless black-hole solutions. We argue
that they can be understood as composite objects. At least one of the
components would have ``negative mass''. We also argue that these
states, being annihilated by all the generators in the supersymmetry
algebra, could also constitute alternative vacua of the supergravity
theory.

\end{quotation}

\begin{flushleft}
CERN--TH/96--214\\
\end{flushleft}

\newpage

\pagestyle{plain}


\section{MASSIVE SUPERSYMMETRIC BLACK HOLES}


\subsection{INTRODUCTION}
\label{sec-intro}

Some of the most interesting classical solutions of string theory are
extreme black holes. Black holes (BHs) are the closest relatives of
massive point particles in gravitational physics. The masses of extreme
BHs are completely determined by their charges. They are the extreme
limit of regular charged BHs with two horizons (an event horizon and a
Cauchy horizon) that coincide in this limit. In some cases they are
regular BHs themselves and in other cases they are not, but in all
cases, they become singular beyond that limit.

Part of the interest in extreme BHs is based on the fact that in many
cases the extreme limit coincides with the limit in which some
supersymmetry is restored. Configurations with unbroken supersymmetry
enjoy very special properties: stability, non-renormalization theorems
etc. and the states of the quantum theory whose field they describe
are usually referred to as BPS states.

There are, though, some unsatisfactory aspects in this correspondence
between extreme and supersymmetric BHs. To start with, the same metric
can correspond to different solutions of the same supergravity theory
(or even different supergravity theories). These solutions, usually
called {\it embeddings} do not always have the same number of unbroken
supersymmetries. Perhaps the best-known example of this situation is
the extreme string dilaton BH ($a=1$) Ref.~\cite{kn:G,kn:GM,kn:GHS}.
When the vector field present in the solution is identified (as in
Ref.~\cite{kn:GHS}) as a Yang-Mills vector field belonging to a matter
supermultiplet, the corresponding embedding is not supersymmetric.
However, if the vector field is identified (as in
Ref.~\cite{kn:KLOPP}) with a graviphoton (one of the six of $N=4$
supergravity) the embedding has two out of four supersymmetries
unbroken. Since the metric of both embeddings is evidently the same,
it is hard to understand why in one case there is unbroken
supersymmetry and in the other case there is not.

Our aim is to clarify this problem and we will give a partial
solution. To this end we will study the embeddings of dilaton BHs
simultaneously in $N=4$ and $N=8$ supergravity (low-energy heterotic
and type~II superstring effective actions). We will find that all
embeddings of the same dilaton BH metric in $N=8$ supergravity are
related by U~duality transformations ({\it i.e.}~there is a unique
embedding modulo U~duality) and have the same number of unbroken
supersymmetries.  However, the $N=4$ embeddings do not have the same
number of unbroken supersymmetries, the reason being that they are not
always related by a T~or S~duality transformation: the missing
U~duality transformation is not in the theory anymore.

There is, though, an exception to this situation for the
Reissner-Nordstr\"om BH: there is a class of embeddings in $N=8$ which
have no unbroken supersymmetries at all. One could, then, imagine a
higher supergravity theory in which all the embeddings of the
Reissner-Nordstr\"om BH where equivalent. A possible scenario in which
this theory has a twelve-dimensional origin has been proposed
\cite{kn:KhO2}. Similar theories have also been proposed on different
grounds (and with different signatures) \cite{kn:H,kn:V}.

Now let us outline the procedure followed. First we obtain the action
and supersymmetry transformation rules of ten-dimensional type~IIA
supergravity by dimensional reduction from $N=1,d=11$ supergravity.
Setting the NS-NS fields to zero one gets $N=1,d=10$ supergravity. Any
solution of the latter is automatically a solution of the former and
this framework allows us to analyze simultaneously the unbroken
supersymmetries of the same metric embedded in both theories. We
further dimensionally reduce to four dimensions, getting $N=4$
supergravity coupled to six vector multiplets. 

Then, in four dimensions we study which dilaton BH metrics can be
embedded in $N=4(+6V)$ supergravity. For the cases in which this is
possible ($a=\sqrt{3},1,1/\sqrt{3},0$), we rewrite the solutions in
ten-dimensional notation using our previous results and study their
unbroken supersymmetries in ten dimensions.

In Section~\ref{sec-dilbh} we present the extreme dilaton BH solutions
and set up the problem of their embedding in $N=4(+6V)$ supergravity.
In Section~\ref{sec-n8} we give the results and comment on them. In
Section~\ref{sec-exception} we deal with the non-supersymmetric
embedding of the Reissner-Nordstr\"om BH.

This first half of the talk is mostly based on the work in
Refs.~\cite{kn:KhO1,kn:KhO2}.


\subsection{THE EMBEDDING OF DILATON BHS IN $N=8$ SUPERGRAVITY}
\label{sec-dilbh}

Dilaton BHs are BH-type solution of the following action:

\begin{equation}
S^{(a)}= {\textstyle\frac{1}{2}}\int d^{4}x\ \sqrt{|g|}
\left[-R -2(\partial\varphi)^{2}
+{\textstyle\frac{1}{2}}e^{-2a\varphi}F^{2}\right]\, .
\end{equation}

\noindent that depends on the real positive constant $a$ and will be 
referred to as ``$a$-model''.  The scalar $\varphi$ is in general
different from the string dilaton which we denote by $\phi$. The
corresponding equations of motion are

\begin{equation}
\left.
\begin{array}{rcl}
G_{\alpha\beta} +2 T_{\alpha\beta}^{\varphi} -e^{-2a\varphi}
T_{\alpha\beta}^{A} & = & 0\, ,
\\
& &
\\
\nabla^{2}\varphi -{\textstyle\frac{a}{4}}e^{-2a\varphi}F^{2} & = & 0\, ,
\\
& &
\\
\nabla_{\mu}\left(e^{-2a\varphi} F^{\mu\alpha}  \right) & = & 0\, ,
\end{array}
\right\}
\label{eq:aeqmo}
\end{equation}

\noindent where $T^{\varphi}_{\alpha\beta}$ and $T^{A}_{\alpha\beta}$
are the energy-momentum tensors of the scalar field $\varphi$ and the
vector field $A_{\mu}$ respectively and $G_{\alpha\beta}$ is the
Einstein tensor.  These equations admit extreme BH \cite{kn:HW} and
multi-BH solutions \cite{kn:S1,kn:O2} for all $a$.  The purely
electric extreme multi-BH solutions are

\begin{equation}
\left\{
\begin{array}{rcl}
ds^{2} & = & V^{-\frac{2}{1+a^{2}}} dt^{2}
-V^{+\frac{2}{1+a^{2}}}d\vec{x}^{2}\, ,
\\
& &
\\
e^{\varphi} & = & V^{-\frac{a}{1+a^{2}}}\, ,
\\
& &
\\
F_{t\underline{i}} & = & -n\ \sqrt{\frac{2}{1+a^{2}}}\
\partial_{\underline{i}}V^{-1}\, ,
\end{array}
\right.
\label{eq:mbhasolutions}
\end{equation}

\noindent where $V(\vec{x})$ is a harmonic function in three-dimensional
Euclidean space ($\partial_{\underline{i}} \partial_{\underline{i}}V
=0$) and $n=\pm 1$ gives the sign of the electric charges.  The
equations of motion of the $a$-model are invariant under the discrete
electric-magnetic duality transformation

\begin{equation}
F^{\prime} =  e^{-2a\varphi} F\, ,
\hspace{1cm}
\varphi^{\prime}  =  -\varphi\, ,
\label{eq:amemdual}
\end{equation}

\noindent and, therefore, a purely magnetic multi-BH solution
always exists for any $a$.  Furthermore, for the special values $a=0$
and $a=1$ dyonic solutions also exist \cite{kn:G,kn:GM,kn:KLOPP}, but
we will need them here.

All these extreme solutions admit Killing spinors if one uses the
appropriate definition of ``gravitino'' and ``dilatino'' supersymmetry
transformation rules \cite{kn:GKLTT}. These rules coincide with the
supersymmetry rules of known supergravity theories only in some cases,
but it is not known exactly in which ones and for which values of $a$.
It was argued in Ref.~\cite{kn:HT} that the cases
$a=1/\sqrt{3},1,\sqrt{3},0$, should appear in different consistent
truncations of maximal $N=8$ supergravity.

Our first goal is to find for which values of $a$ the dilaton-BH
solutions can be embedded in $N=4(+6V)$ supergravity (low-energy
effective heterotic string theory or type~IIA theory, in the sense
explained before). This is the same as solving the following problem:
given the equations of motion of $N=4(+6)$ supergravity

\begin{eqnarray}
G_{\alpha\beta} +2 T_{\alpha\beta}^{\phi}
+{\textstyle\frac{9}{4}} T_{\alpha\beta}^{B}
& &
\nonumber \\
& &
\nonumber \\
-{\textstyle\frac{1}{4}} \left[\partial_{\alpha}G_{mn}
\partial_{\beta} G^{mn} -{\textstyle\frac{1}{2}}
g_{\alpha\beta} \partial_{\mu}G_{mn} \partial^{\mu}G^{mn}
\right]
& &
\nonumber \\
& &
\nonumber \\
-{\textstyle\frac{1}{4}} G^{mn} G^{pq} \left[\partial_{\alpha}B_{mp}
\partial_{\beta} B_{nq} -{\textstyle\frac{1}{2}}
g_{\alpha\beta} \partial_{\mu}B_{mp} \partial^{\mu}B_{nq}
\right]
& &
\nonumber \\
& &
\nonumber \\
+{\textstyle\frac{1}{2}} G_{mn} \left[F^{(1)m}{}_{\alpha}{}^{\mu}
F^{(1)m}{}_{\beta\mu} -{\textstyle\frac{1}{4}}
g_{\alpha\beta}F^{(1)m}{}_{\mu\nu} F^{(1)m\mu\nu} \right]
& &
\nonumber \\
& &
\nonumber \\
+{\textstyle\frac{1}{2}} G^{mn} \left[{\cal F}_{m\alpha}{}^{\mu}
{\cal F}_{n\beta\mu} -{\textstyle\frac{1}{4}}\tilde{g}_{\alpha\beta}
{\cal F}_{m\mu\nu} {\cal F}_{n}{}^{\mu\nu} \right]
& = & 0\, , \\
& &
\nonumber \\
\nabla^{2}\phi +{\textstyle\frac{3}{4}}e^{-4\phi}H^{2}
+{\textstyle\frac{1}{8}}e^{-2\phi}\left[G_{mn}F^{(1)m} F^{(1)n}
+G^{mn}{\cal F}_{m} {\cal F}_{n}\right]
& = & 0\, ,
\\
& &
\nonumber \\
\nabla^{2}G^{rs} -G^{m(r}G^{s)n}G^{pq}\left[\partial G_{mp}
\partial G_{nq} +\partial B_{mp} \partial B_{nq} \right]
& &
\nonumber \\
& &
\nonumber \\
+{\textstyle\frac{1}{2}} e^{-2\phi}\left[F^{(1)r}F^{(1)s}
-G^{m(r}G^{s)n}{\cal F}_{m} {\cal F}_{n}\right]
& = & 0\, ,
\\
& &
\nonumber \\
\nabla_{\mu}\left(G^{nr}G^{qs}\partial^{\mu}B_{nq} \right)
+e^{-2\phi}{\cal F}_{m}G^{m[s}F^{(1)r]}
& = & 0
\\
& &
\nonumber \\
\nabla_{\mu}\left( e^{-2\phi} G_{mn} F^{(1)n\mu\alpha}\right)
& = & 0\, ,
\\
& &
\nonumber \\
\nabla_{\mu}\left( e^{-2\phi}G^{mn}{\cal
F}_{n}{}^{\mu\alpha}\right)
& = & 0\, ,
\\
& &
\nonumber \\
\nabla_{\mu}\left( e^{-4\phi}H^{\mu\alpha\beta}\right)
& = & 0\, ,
\end{eqnarray}

\noindent where $T^{\phi}_{\alpha\beta}$ and $T^{B}_{\alpha\beta}$ are 
the energy-momentum tensors of $\phi$ and $B_{\alpha\beta}$, in how
many ways can they be reduced to those above of the $a$-model? One
also has to take into account the Bianchi identities

\begin{equation}
\begin{array}{rclrcl}
\partial F^{(1)m}
&
=
&
0\, ,
&
\partial H
&
=
&
{\textstyle\frac{1}{2}}F^{(1)m}F^{(2)}{}_{m}\, ,
\\
& & & & & \\
\partial F^{(2)}{}_{m}
&
=
&
0\, .
& & & \\
\end{array}
\label{eq:bianchis}
\end{equation}

It is clear that, to reduce these equations to the equations of the
$a$-model, which only has one scalar and one vector field one has to
identify the different vector field strengths either with $F$ or with
${}^{\star}F$ or with linear combinations of both. Something analogous
happens with the scalars. The result is that only in the cases
$a=\sqrt{3},1,1/\sqrt{3},0$ this can be done and it can be done in a
few ways and under certain constraints: $F$ has to be purely electric
(or magnetic). This is explained in the next section.


\subsection{$N=8$ SUPERSYMMETRIC BLACK HOLES}
\label{sec-n8}

Our results are collected in Table~\ref{tab-embeddings}.  When the
$N=4(+6V)$ fields of the top row take the values given in the
following rows, in terms of $\varphi$ and $F$, where $F$ is either
purely electric or purely magnetic, then the $N=4(+6V)$ equations of
motion reduce to those of the $a$-model for the value of $a$ given in
the first column.

These are all the possible embeddings up to S~or T~duality symmetries
of the $N=4(+6V)$ theory. Once the embeddings are known, it is easy to
re-express the solutions as solutions of $N=1,d=10$ and $N=2A,d=10$
supergravity and study the unbroken supersymmetries. The
supersymmetries of the $N=1$ theory correspond to one chiral sector of
the $N=2A$ theory (conventionally, the positive chirality sector) and
are given in last column under $n_{+}$. The unbroken supersymmetries
of the other chiral sector, which are in the $N=2A$ but not in the
$N=1$ theory are under $n_{-}$\footnote{An $N=1$ theory with negative
  chirality can also be constructed and fits in the negative chirality
  sector of the $N+2A$ theory as explained in Ref.~\cite{kn:KhO1}.}.
For each value of $a$ allowed, the total unbroken supersymmetry in the
$N=2A$ theory ($n_{+}+n_{-}$) is always the same, but $n_{\pm}$ are in
general different for each $N=1$-inequivalent embedding, as explained
in the introduction.

\begin{table}
\begin{center}
\begin{tabular}{||c||c|c|c|c|c|c|c||c||}
\hline\hline
& & & & & & & & \\
$a$ & $\phi$ & $\rho_{1}$ & $\rho_{2}$ &
$F^{(1)1}$ & $F^{(2)}{}_{1}$ & $F^{(1)2}$ & $F^{(2)}{}_{2}$ &
$(n_{+},n_{-})$ \\
\hline\hline
& & & & & & & & \\
$\sqrt{3}$ & $\frac{1}{\sqrt{3}}\varphi$ &
$-\frac{2}{\sqrt{3}}\varphi$ & $0$ &
$\sqrt{2}F$ & $0$ & $0$ & $0$ & $(\frac{1}{2},\frac{1}{2})$
\\
\hline\hline
& & & & & & & & \\
$1$ & $\varphi$ & $0$ & $0$ & $F$ & $-F$ & $0$ & $0$ & $(\frac{1}{2},0)$
\\
\cline{2-9}
& & & & & & & & \\
 & $\varphi$ & $0$ & $0$ & $F$ & $+F$ & $0$ & $0$ & $(0,\frac{1}{2})$
\\
\cline{2-9}
& & & & & & & & \\
 & $0$ & $-\varphi$ & $\varphi$ & $F$ & $0$ & $e^{-2\varphi}{}^{\star}F$
& $0$ & $(\frac{1}{4},\frac{1}{4})$
\\
\hline\hline
& & & & & & & & \\
$\frac{1}{\sqrt{3}}$ & $-\frac{1}{3}\varphi$ & $-\frac{2}{3}\varphi$ &
$0$ & $\sqrt{\frac{2}{3}}F$ & $0$ &
$\sqrt{\frac{2}{3}}e^{2\phi}{}^{\star}F$
& $-\sqrt{\frac{2}{3}}e^{2\phi}{}^{\star}F$ & $(\frac{1}{4},0)$
\\
\cline{2-9}
& & & & & & & & \\
 & $-\frac{1}{3}\varphi$ & $-\frac{2}{3}\varphi$ &
$0$ & $\sqrt{\frac{2}{3}}F$ & $0$ &
$\sqrt{\frac{2}{3}}e^{2\phi}{}^{\star}F$
& $+\sqrt{\frac{2}{3}}e^{2\phi}{}^{\star}F$ & $(0,\frac{1}{4})$
\\
\hline\hline
& & & & & & & & \\
$0$ & $0$ & $0$ & $0$ & $\frac{1}{\sqrt{2}}F$ &
$-\frac{1}{\sqrt{2}}F$ & $\frac{1}{\sqrt{2}}{}^{\star}F$ &
$-\frac{1}{\sqrt{2}}{}^{\star}F$ & $(\frac{1}{4},0)$
\\
\cline{2-9}
& & & & & & & & \\
 & $0$ & $0$ & $0$ & $\frac{1}{\sqrt{2}}F$ &
$+\frac{1}{\sqrt{2}}F$ & $\frac{1}{\sqrt{2}}{}^{\star}F$ &
$+\frac{1}{\sqrt{2}}{}^{\star}F$ & $(0,\frac{1}{4})$
\\
\cline{2-9}
& & & & & & & & \\
 & $0$ & $0$ & $0$ & $F\pm {}^{\star}F$ & $0$ & $0$ & $0$ & $(0,0)$
\\
\hline\hline
\end{tabular}
\end{center}
\caption[a]{Table of embeddings and supersymmetries of dilaton BHs.}
\label{tab-embeddings}
\end{table}


\subsection{THE EXCEPTION TO THE RULE}
\label{sec-exception}

The only exception corresponds to the last row in
Table~\ref{tab-embeddings} which is a dyonic embedding of the
Reissner-Nordstr\"om BH metric with no supersymmetries and,
consequently, inequivalent under U~duality to the other embeddings.
The corresponding ten-dimensional configuration is purely
gravitational 

\begin{equation}
\left\{
\begin{array}{rcl}
d\hat{s}^{2} & = & V^{-2}dt^{2} -V^{2}d\vec{x}^{2}
-\left[ dx^{\underline{4}}
+\sqrt{2}\ n\ \left( V^{-1}dt \pm V_{\underline{i}}
dx^{\underline{i}}\right) \right]^{2}\\
& & \\
& &
-dx^{\underline{I}} dx^{\underline{I}}\, ,
\hspace{1cm} I=5,\ldots,9\, .\\
& & \\
\hat{B} & = & \hat{\phi} = 0\, ,\\
\end{array}
\right.
\end{equation}

\noindent where the $V_{\underline{i}}$'s are functions satisfying

\begin{equation}
\partial_{[\underline{i}}V_{\underline{j}]} =
{\textstyle\frac{1}{2}}\epsilon_{ijk}\partial_{\underline{k}}V\, ,
\end{equation}

\noindent and this makes it very easy to check that $n_{+}=n_{-}=0$.

It is somewhat surprising to find that the solution to our problem
(namely embedding the $N=4(+6V)$ solutions into maximal supergravity
so all the embeddings become equivalent) does not work in this case.
A possibility is that there exists a higher supergravity theory that
includes $N=8$ as a consistent truncation and which has a bigger
duality group which makes absolutely all embeddings of the same
metric equivalent. A possible scenario \cite{kn:KhO2} is related to
the presence of an $SL(2,\Z)$ duality group in the $N=2B$ supergravity
theory whose origin could be twelve-dimensional \cite{kn:BJO}.


\section{MASSLESS BLACK HOLES}
\label{sec-massless}

This second half of the talk is based on Ref.~\cite{kn:O1}.

In General Relativity the ADM mass $m$ of an asymptotically flat space,
which describes an isolated system, is the total energy of that system.
Therefore, zero ADM mass should mean empty ({\it i.e.}~flat) space.
Indeed, when the ADM mass $m$ of the Schwarzschild metric goes to zero

\begin{equation}
\left(1 -\frac{2m}{r} \right) dt^{2}
-\left(1 -\frac{2m}{r} \right)^{-1} dr^{2} -r^{2}d\Omega^{2}\hspace{.5cm}
\stackrel{m\rightarrow 0}{\longrightarrow}\hspace{.5cm}
dt^{2} -d\vec{x}^{2}\, ,
\end{equation}

\noindent it approaches Minkowski's. However, things are different
for the Reissner-Nordstr\"om (RN) metric:

\begin{eqnarray}
\left(1 -\frac{2m}{r} +\frac{q^{2}}{r^{2}}\right) dt^{2}
-\left(1 -\frac{2m}{r} +\frac{q^{2}}{r^{2}}\right)^{-1} dr^{2}
-r^{2}d\Omega^{2} \hspace{-4cm}
& & \nonumber \\
& &  \nonumber \\
& & \hspace{-2.5cm} \stackrel{m\rightarrow 0}{\longrightarrow} 
\left(1 +\frac{q^{2}}{r^{2}}\right) dt^{2}
-\left(1 +\frac{q^{2}}{r^{2}}\right)^{-1} dr^{2}
-r^{2}d\Omega^{2}\, .
\label{eq:RNm0}
\end{eqnarray}

\noindent This metric describes a charged massless system and has a
naked singularity at $r=0$\footnote{The singularity would be covered
  by a regular horizon if the sign of $q^{2}$ was reversed or the
  electric charge $q$ was Wick-rotated. This solution, describing a
  massless genuine black hole would not be a solution of the
  Einstein-Maxwell theory anymore, but it would be a solutions of the
  Einstein-anti-Maxwell theory considered by Gibbons and Rasheed in
  Ref.~\cite{kn:GR}. We will not consider this kind of theories
  here.}. If we were considering macroscopic objects we could invoke
Penrose's cosmic censorship hypothesis (which is an statement on the
{\it evolution} of gravitating systems): the above metric will never
be the result of the gravitational collapse of energetically
well-behaved matter.

In fact, we know that the presence of an electromagnetic field gives a
positive contribution to the total energy.  There must be a negative
contribution to compensate it which would already be present before
gravitational collapse took place and would violate the energy
conditions of the cosmic censorship hypothesis.  However, in GR it is
impossible to give a rigorous definition of a local energy density and,
if we are only given the metric describing the final state, we cannot
say where the negative energy is and what its origin is.

The cosmic censorship hypothesis, though, does not apply to miscroscopic
objects which are not the result of gravitational collapse.  In fact, we
do not know how bad the effect of the naked singularity of a microscopic
object is, since, anyway, this metric would only be the (far field)
low-energy effective description of that object.

Here is where supersymmetry comes to our rescue: the mass of all the
quantum states of a supersymmetric theory satisfy a so-called
Bogmol'nyi bound \cite{kn:WO}.  In particular, we could consider the
RN metric as a solution of $N=2$ supergravity (which is just
Einstein-Maxwell's theory when all the fermions are set to zero)
describing the field of an associated quantum state of mass $m$ and
charge $q$.  The B-bound in this theory is simply

\begin{equation}
m \geq |q+ ip|\, ,
\end{equation}

\noindent where $p$ is the magnetic charge. In these conditions we
could say that the singular metric (\ref{eq:RNm0}) does not describe
the gravitational field of any state of the theory and we could
discard it. Massless charged states of pure $N=2$ supergravity seem to
be excluded and also supersymmetry seems to act as a cosmic censor at
the microscopic level \cite{kn:KLOPP} (but see later).

For higher $N$ extended supergravities the B-bound depends on the
moduli $\phi_{0}$: $m\geq |Z(\phi_{0},q,p)|$ and it was argued in
\cite{kn:HT2} that for specific values of the moduli, the form
$Z(\phi_{0},q,p)$ could be singular and become zero for certain values
of the charges. States saturating the bound would be massless, and the
appearance of these new massless states would be the signal of
symmetry enhancement\footnote{One has to be a bit careful in the
interpretation of this result, though. The B-bound formula is deduced
from the supersymmetry algebra assuming that one is dealing with a {\it
massive} representation and then going to the rest frame
$p^{\mu}=(m,\vec{0})$. The analysis is not valid anymore at the singular
point in moduli space where the representation becomes {\it massless}.
At the critical point, a different analysis is required for that
representation.}.

Soon after this proposal was made, massless supersymmetric ({\it
i.e.}~B-bound-saturating) black holes were found in
Refs.~\cite{kn:Be,kn:KL1,kn:KL2}.  The canonical metric of the one we
are going to concentrate in is

\begin{equation}
\label{eq:metric}
ds^{2} = \left( 1- \frac{D^{2}}{r^{2}} \right)^{-\frac{1}{2}} dt^{2}
-\left( 1- \frac{D^{2}}{r^{2}} \right)^{\frac{1}{2}}d\vec{x}^{2}\, .
\end{equation}

Objects of this kind cannot be excluded using our previous arguments:
they are allowed by supersymmetry and they are also exact solutions of
string theory. Understanding the nature of these objects is not only a
challenge but a must to be able to exclude them or to rightly include
them in the theory. Our purpose here will be to elaborate a model for
them. We start  by reviewing some of their main features:

\begin{enumerate}

\item The metric is singular when $r=|D|$.  The singularity is a
      curvature singularity (observe that the area of spheres of radius
      $r$, $4\pi \left( 1- \frac{D^{2}}{r^{2}}
      \right)^{\frac{1}{2}}r^{2}$, goes to zero in that limit).

\item This metric does not seem to be the extreme limit of any
      regular charged black-hole metric.

\item The expansion of the $g_{tt}$  component of the metric far away
      from the singularity, where the gravitational field is weak is

\begin{equation}
g_{tt}= 1 +\frac{D^{2}}{2}\frac{1}{r^{2}}
+\frac{3D^{4}}{8}\frac{1}{r^{4}}
+{\cal O}\left( \frac{1}{r^{6}} \right)\, .
\label{eq:gtt}
\end{equation}

The coefficient of the $\frac{1}{r}$ term is, by definition, $-2m$ and
so the ADM mass of these objects is zero.  In this limit, $g_{tt}\sim
1+2\Phi$, where $\Phi$ is the Newtonian gravitational potential.
Therefore,

\begin{equation}
\Phi\sim
\frac{D^{2}}{4}\frac{1}{r^{2}}
+\frac{3D^{4}}{16}\frac{1}{r^{4}}\, ,
\label{eq:Phi}
\end{equation}

\noindent and has weakly repulsive character when acting
on usual test particles \cite{kn:KL1}.

\item They do not seem to move at the speed of light (or to move at
      all). In GR, an object moving at the speed of light would be
      represented by a gravitational wave. But this object is not a wave
      because it does not admit any light-like Killing vector to start
      with. On the other hand, usual objects
      with zero rest mass moving at the speed of light have positive
      total energy and non-zero  three-momentum, unlike massless BHs.
      Actually, the  whole ADM four-momentum of these objects
      vanishes.

\item When they are rightly embedded in a supergravity theory, they have
      half of $N=2$ or $N=4$ supersymmetries unbroken and the
      low-energy solutions describing them are also exact solutions of
      string theory.

\item The last two properties imply stability: they cannot decay into
      anything else because they have the minimal mass allowed by
      supersymmetry (they saturate a B-bound) and the minimal
      mass allowed in Physics (zero).

\item There are no known metrics describing an arbitrary number of
      these objects in static equilibrium.

\end{enumerate}

This last point deserves some further explanation.  In the
Einstein-Maxwell theory (which is the bosonic sector of $N=2$
supergravity) there is a family of solutions known as the
Majumdar-Papapetrou (MP) solutions which are given in terms of a real
function $V(\vec{x})$ harmonic in three-dimensional Euclidean space
$\partial_{\underline{i}} \partial_{\underline{i}} V=0$

\begin{equation}
\left\{
\begin{array}{rcl}
ds^{2} & = & V^{-2} dt^{2} -V^{2}d\vec{x}^{2}\, ,
\\
& &
\\
F_{t\underline{i}} & = & \pm\ \sqrt{2}\
\partial_{\underline{i}}V^{-1}\, ,
\end{array}
\right.
\end{equation}

$V$ can be chosen at will.  All these solutions admit Killing spinors
\cite{kn:GHu} and for choices that make the metric asymptotically flat,
they are supersymmetric.  The most interesting choice for $V$ is with
point-like singularities:

\begin{equation}
V(\vec{x})= 1 +\sum_{i}\frac{m_{i}}{|\vec{x}-\vec{x}_{i}|}\, .
\end{equation}

When all the constants $m_{i}>0$, the metric describes a set of extreme
RN black holes in static equilibrium \cite{kn:HaHa} with charges
$\pm m_{i}$. The (total) ADM mass is $m=\sum_{i}m_{i}$. One is tempted
to say that the $ith$ BH has mass $m_{i}$, but there is no way to prove
this, no matter how logical this conclusion seems to be, and the only
well-defined mass in this system is $m$. However, thinking in terms of
the individual masses can be physically helpful.

It is then possible to have supersymmetric solutions with zero ADM mass
if we choose the constants $m_{i}$ (now positive and negative) such that
$\sum_{i}m_{i}=0$.  For instance, we can have pairs of objects with
opposite values of $m_{i}$.  Each pair, if it was isolated, would be
massless and we could say that the solution describes many of these
massless pairs in equilibrium (even though the mass of each pair is not
well defined).  The constituents of these pairs would annihilate each
other if placed at the same point and, therefore, this kind of solutions
are quantum-mechanically unstable.

The only solutions known so far describing many massless BHs are of this
kind.  They never describe many objects of the type we are considering,
but pairs of objects.  This observation, together with the fact that
elementary massless objects move at the speed of light, will prove
important in what follows, because they are against an interpretation of
massless BHs as elementary objects.  Furthermore, the fact that there is
a gravitational field although the total mass is zero suggests that the
source of the gravitational field is an object with complex structure
whose far field is approximately described by the above metric. It
will be useful to establish an electrostatic analogy.


\subsection{AN ELECTROSTATIC MODEL}
\label{sec-electrostatic}

Let us consider the electrostatic potential $\phi$ of a spatially
confined charge distribution. If we perform a multi-pole expansion and
find that the $1/r$ term is absent, we would immediately conclude that
the total charge is zero. However, if the field is not trivial, we must
also conclude that there is some charge, in fact, as much positive as
negative charge. This system resembles our massless BH but there is
also a big difference: the field cannot be spherically symmetric (the
only spherically symmetric term is the $1/r$ one).

To get closer to the massless BH gravitational field one has to consider
charge distributions that fill the whole space.  This is not surprising
because of the non-Abelian nature of the gravitational field: the
gravitational field fills the whole space and is the source of
gravitational field.  Thus, let us consider two different, spherically
symmetric, charge distributions $\rho^{(+)}(r)>0$ and $\rho^{(-)}(r)<0$
such that

\begin{equation}
\int dV \rho^{(+)} = Q >0\, ,
\hspace{1cm}
\int dV \rho^{(-)} = -Q\, .
\end{equation}

\noindent so the total charge density is $\rho
(r)=\rho^{(+)}+\rho^{(-)}\neq 0$ is also spherically symmetric and the
total charge is zero.  At a distance $r$ of the origin, the
electrostatic potential behaves as

\begin{equation}
\phi(r)\sim Q(r)/r\, ,
\end{equation}

\noindent where $Q(r)$ is the charge contained in a sphere of radius $r$

\begin{equation}
Q(r)= \int_{r^{\prime}\leq r}dV \rho (r^{\prime})\, .
\end{equation}

Now, if the charge densities are such that, for large $r$, $Q(r)\sim
1/r$, then $\phi\sim 1/r^{2}$, which is the large $r$ behavior of the
Newtonian potential for the massless BHs.

Then, we have been able to generate a spherically symmetric
electrostatic field with zero total charge but at the expense of having
positive and negative charge densities in the whole space.


\subsection{A MODEL FOR MASSLESS BHS}
\label{sec-model}

The electrostatic analogy together with the previous discussion suggest
a model for a massless BH: a massless BH would be a composite object,
the components being two ``energy distributions'' that separately would
correspond to equal but opposite total masses. (In GR a point-like
object gives an energy distribution that fills the whole space.)

But this system is unstable  classically (the components repel
each other) and quantum-mechanically (they could annihilate each other).
Furthermore, since the components repel each other, there would be a
non-vanishing interaction energy and the total system would not be
massless as we naively expected.

To solve these problems we need to have additional interactions so the
total force felt by each component vanishes and the interaction energy
also vanishes.  This behavior is common when there is unbroken
supersymmetry and this is essentially what we have in the massless MP
solutions.  We need, however, more than one additional kind of charge to
overcome the possibility of annihilation between pairs with opposite
amount of the same kind of charge that we found in the massless MP
solutions.  But, if the components carry different kinds of charge, they
would not interact!  This new problem is solved by the presence of
scalar fields which interact with different kinds of charges.

Then, massless BHs of the type we are considering could be found in
supergravity theories with $N>2$ (or $N=2$ with matter multiplets) and
they could be made out of pairs with opposite ``masses'' and unbroken
supersymmetry (in multi-BH solutions).

It is not difficult to check this hypothesis.  The multi-BH solution we
need was found in Refs.~\cite{kn:CY,kn:CT} and later rediscovered in
Ref.~\cite{kn:R} in the framework of the theory described by the
following simple action

\begin{eqnarray}
S & = & \int dx^{4}\ \sqrt{-g} \left\{ -R
-2\left[ (\partial \phi)^{2} +(\partial \sigma)^{2}
+(\partial \rho)^{2}\right] \right.
\nonumber \\
& & \nonumber \\
& &
+{\textstyle\frac{1}{4}} e^{-2\phi}
\left[
e^{-2(\sigma+\rho)} (F^{(1)1})^{2}
+e^{-2(\sigma-\rho)} (F^{(1)2})^{2}
\right.
\nonumber \\
& & \nonumber \\
& &
\left.
\left.
+e^{2(\sigma+\rho)} (F^{(2)}{}_{1})^{2}
+e^{2(\sigma-\rho)} (F^{(2)}{}_{2})^{2} \right] \right\}\, ,
\end{eqnarray}

\noindent which is a truncation of the low-energy effective action
of the heterotic string \cite{kn:MS} ($N=4$ supergravity plus 22
vector multiplets).

The solution is given in terms of four independent harmonic functions
$H^{(1)},K^{(1)},H^{(2)},K^{(2)}$ ($\partial_{\underline{i}}
\partial_{\underline{i}} H =\partial_{\underline{i}}
\partial_{\underline{i}} K=0\, ,\,\,\, i=1,2,3$)

\begin{eqnarray}
ds^{2} & = & U^{-\frac{1}{2}} dt^{2} -U^{\frac{1}{2}} d\vec{x}^{2}\, ,
\hspace{1cm} U= H^{(1)}K^{(1)}H^{(2)}K^{(2)}\, ,
\nonumber \\
& & \nonumber \\
e^{-4\phi} & = & \frac{H^{(1)}H^{(2)}}{K^{(1)}K^{(2)}}\, ,
\hspace{1cm}
e^{-4\sigma} =\frac{H^{(1)}K^{(2)}}{H^{(2)}K^{(1)}}\, ,
\hspace{1cm}
e^{-4\rho} =\frac{H^{(1)}K^{(1)}}{H^{(2)}K^{(2)}}\, ,
\nonumber \\
& & \nonumber \\
F^{(a)1}{}_{t\underline{i}} & = &
c^{(a)} \partial_{\underline{i}}\frac{1}{H^{(1)}}\, ,
\hspace{.5cm}
\tilde{F}^{(a)2}{}_{t\underline{i}} =
d^{(a)} \partial_{\underline{i}}\frac{1}{K^{(1)}}\, ,
\hspace{.5cm}
a=1,2\, .
\end{eqnarray}

\noindent where $(c^{(a)})^{2}=(d^{(a)})^{2}=1$ and

\begin{equation}
\tilde{F}^{(1)2} =e^{-2(\phi+\sigma-\rho)}
{}^{\star}F^{(1)2}\, ,
\hspace{1cm}
\tilde{F}^{(2)}{}_{2} =e^{-2(\phi-\sigma+\rho)}
{}^{\star}F^{(2)}{}_{2}\, ,
\end{equation}

\noindent and ${}^{\star}F$ is the Hodge dual of $F$.  Usually, the
$H$'s and $K$'s are chosen to be strictly positive, that is, all the
constants in

\begin{equation}
H^{(a)}=1 +\sum_{n}\frac{q_{n}^{(a)}}{|\vec{x}-\vec{x}_{n}|}\, ,
\hspace{1cm}
K^{(a)}=1 +\sum_{n}\frac{p^{(a)}_{n}}{|\vec{x}-\vec{x}_{n}|}\, ,
\end{equation}

\noindent are non-negative to avoid singularities in the metric, but we
are after solutions with some negative $q$'s or $p$'s.  Now, following
Ref.~\cite{kn:R}, we consider solutions of the form

\begin{equation}
H^{(a)} = 1 +\frac{q_{a}}{|\vec{x}-\vec{x}_{1}|}\, ,
\hspace{.5cm}
K^{(a)} = 1 +\frac{p_{a}}{|\vec{x}-\vec{x}_{2}|}\, .
\end{equation}

When all the $q$'s and $p$'s but one vanish, the solution is an
$a=\sqrt{3}$ extreme dilaton BH if the non-vanishing constant is
positive.  Then, if several constants are positive, one can consider
that the above solutions describes as many $a=\sqrt{3}$ BHs in
equilibrium.  The ADM mass of the system is $m ={\textstyle\frac{1}{4}}
(q_{1}+p_{1}+q_{2}+p_{2})$ and would be positive.  When the coordinates
of all the BHs coincide at one point (or we look to the system from far
away so the distances between black holes are negligible) one gets
$a=1,1/\sqrt{3},0$ extreme dilaton BHs which can be seen as a bound
state with zero binding energy of elementary $a=\sqrt{3}$ BHs.

If we now take, for instance $q^{(1)}=-q^{2)}=D$ and $p^{(1)}=p^{(2)}=0$
(that is, an $a=\sqrt{3}$ BH-``anti-BH'' pair\footnote{The mass of each
individual component would be $D/4$ and $-D/4$ if we could define the
mass of each component, which is not possible: the only mass that can be
rigorously defined is the total ADM mass of the whole space-time,
$m$.}, or {\it dihole}), we get the following massless metric

\begin{equation}
ds^{2}= \left(1+\frac{D}{r_{1}} \right)^{1/2} \left(1-\frac{D}{r_{2}}
\right)^{1/2} dt^{2}
-\left(1+\frac{D}{r_{1}} \right)^{-1/2} \left(1-\frac{D}{r_{2}}
\right)^{-1/2}d\vec{x}^{2}\, .
\end{equation}

\noindent which reproduces the original massless BH metric
(\ref{eq:metric}) when $r_{1}=r_{2}=r$.

Following the same reasoning as in Ref.~\cite{kn:R} we can conclude
that the known massless BHs are the effective field of a bound
state of a pair of objects with opposite masses, or {\it dihole}.


\subsection{CONCLUSION}
\label{sec-conclusion}

Massless black holes are objects with very strange properties. In
particular, it is not clear what they are and whether they should be
included in the theory. Their masslessness and the fact that they
saturate at least a B-bound (because they are supersymmetric) mean
together that they saturate {\it all} bounds (and ``anti-bounds'' of
the form $m+|z_{i}|\geq 0$) and that they are in fact annihilated by
all supersymmetry and Lorentz charges. In this respect they are
analogous to Minkowski space which is also massless and is annihilated
by all charges, being a vacuum state of the theory. Perhaps the answer
to the question of whether massless black holes should move at the
speed of light or not can be found on this analogy because nobody ever
asks whether Minkowski's space should move at the speed of light. A
vacuum state interpretation could perhaps be more appropriate for
massless ``black holes''.

On the other hand, although we have not found any inconsistency in the
inclusion of these admittedly rather exotic objects in the theory many
more checks are clearly necessary. A check that has recently been
performed in Ref.~\cite{kn:E} is the study of the pair-production of
these massless black holes. The study is based in a C-metric-type
solution describing two massless black holes with opposite $U(1)$
charges accelerating apart. This happens without any additional
external force. This is what would happen in QED between electrons and
positrons if $e$ was imaginary, leading to vacuum instability under
charged pair production. It is, therefore, not completely surprising
that the action of related instanton describing the pair production of
massless black holes indicates the corresponding instability of the
vacuum. It is, perhaps, more surprising that two massless black holes
with opposite charges seem to repel each other. The reason why this
happens is not yet completely clear and should be investigated.


\end{document}